# ADELIC THEORY OF STOCK MARKET

Zharkov V.M.[a,b]

[a] Natural science institute of Perm state university,
Genkel st. 4, Perm, 614990, Russia
vita@psu.ru}

[b] Perm state university, Bukireva st. 15, Perm, 614990, Russia

## Abstract

The p-adic theory of the stock market is presented. It is shown that the price dynamics is very naturally described by the adelic function. The procedure of derivation of the functional integral formulation of adelic type is derived from microscopic models using generalized supercoherent states.

## 1. Introduction

We live in the high technology world. We use in finance the artificial neural nets, genetic and evolutionary algorithms investigating financial markets. Econophysics is the bright example of a new high technology theory in finance [1]. Today the new scientific concepts penetrate to modern economic theory, for example, the nonlinear dynamics, deterministic chaos, fractals, fuzzy sets, and others – promising us new discoveries, but at the same time, its appearance prompting the revision of an earlier ones. It is shown in this article that there exist the relationship between the Elliott theory and p-adic description of the dynamics of prices in the stock market. It is reasonable to talk about the existence of a new type of waves in form of steps that are absent in the Elliott theory. The new theory of the stock market, describing the ensemble of traders and containing adelic description of price dynamics was developed

## 2. Elliott theory

In nanotechnology, magnetism, in high-temperature superconductivity and in many physical phenomena we have fractal behavior in the experimental data. The peculiarity of this phenomenon is that the behavior of physical quantities that depend on the time or the magnetic field is non-analytic. One-dimensional fractal as a function of time, the magnetic field, temperature is described by curve, non-differentiable nowhere, then there is a function value or its derivative will be discontinuous at any point. In the late 1920's R. Elliott developed the theory of waves assuming some kind of regularity in the stock markets, contrary to popular assumptions about the random nature of price movement. He found that price movements have repetitive cycles, which are associated with the emotions of investors as a result of external influences of news or mass psychology prevailing at the time. Elliott said that the ascending and descending oscillations of mass psychology always manifest themselves in the same repetitive patterns, which he called "waves " [2].

The wave principle posits that collective investor psychology or crowd psychology moves from optimism to pessimism and back again in a natural sequence. These swings create patterns, as evidenced in the price movements of a market at every degree of trend. Elliott's model says that market prices alternate between five waves and three waves at all degrees of trend, as the illustration shows. Within the dominant trend, waves 1, 3, and 5 are "motive" waves, and each motive wave itself subdivides in five waves. Waves 2 and 4 are "corrective" waves, and subdivide in three waves. In a bear market the dominant trend is downward, so the pattern is reversed—five waves down and three up. Motive waves always move with the trend, while corrective waves move against it.

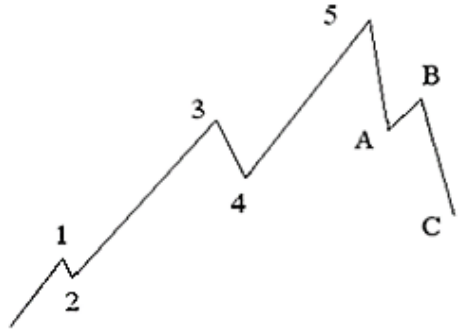

*Fig. 1 shows a fractal of the first level - the curve that is not differentiable at a finite number of points 1, 2, 3, 4, 5, A, B, C.*

The physicist Didier Sornette, professor at the Department of Earth and Space Science and the Institute of Geophysics and Planetary Physics at UCLA. In a paper he co-authored in 1996 ("Stock Market Crashes, Precursors and Replicas") Sornette said, "It is intriguing that the log-periodic structures documented here bear some similarity with the "Elliott waves" of technical analysis. A lot of effort has been developed in finance both by academic and trading institutions and more recently by physicists (using some of their statistical tools developed to deal with complex times series) to analyze past data to get information on the future. The 'Elliott wave' technique is probably the most famous in this field. We speculate that the "Elliott waves", so strongly rooted in the financial analysts' folklore, could be a signature of an underlying critical structure of the stock market"[3].

### 3. How p-adic mathematics appears in the finance

It is a fact but we have never a case of irrational real numbers in everyday life and in scientific experiments. The results of any action we can express only in rational numbers. Of course, there is a common belief that if we measure with greater precision, we can get any number of decimals and interpret the result as a real number. However, this is an idealization, and we must be careful with such statements, therefore, we take as our starting point the field of rational numbers Q. The p-adic analysis and p-adic mathematical physics attract today a great interest. P-adic models have been introduced to the string theory, p-adic quantum theory and p-adic quantum gravity. In the p-adic theory of consciousness were developed. We would like to discuss the applicability of p-adic numbers and adeles to the stock market. Let us give the arguments of appearance of the p-adic numbers in general class of systems. As we have understand for the appearance of such a p-adic formalism it is necessary to apply the functional integral for formulation of systems dynamics, which gives the possibility of a nontrivial change of variables in functional integral from the real valued fields to the p-adic valued fields. This transformation of the fields gives us the new formulation of representation of systems dynamics. As a result we have obtained an effective theory with another set of fields and the different symmetry. As the first step we have used the most evident procedure of introduction the p-adic numbers. We began with systems which had some set of dynamic fields or variables.These variables have some experimental meaning and that is why they have values in the field of rational numbers . Usually the securities have the following values - 10/12, 45/12. The securities mean shares and it is natural that their values fall in the field of rational numbers. In reality it is impossible to obtain the irrational number of investor's share of capital. As a result we have come to the following statement: all the variables which describe the securities are the elements of the rational field [4].

In January 2000 the Commission on the Securities and Exchange Commission give guidance to all major U.S. stock exchanges to transfer all stock quotation systems and systems of registration of transactions with shares and options in the format of the decimal point.

The second step of the construction of any theory is the choice of some method of the final quantities evaluation. We need here a certain procedure for the absolute values evaluation, as well as a procedure of comparison of two numbers. According to Ostrovskii theorem we have two possibilities for completion of rational numbers with module: it can be a real module (the real numbers field) or p-adic module (the p-adic number field). We have an infinite number of p-adic norms which are characterized by a prime number p. At present the real numbers are used by the vast majority of theories, describing the reality, and there is the usual consensus that the real numbers are the main elements for presenting the reality. We intend here to show that p-adic numbers are more suitable for the purposes of description of the financial market price dynamics. Let us define some basic notation. P-denote a prime number. An arbitrary rational number x can be written in the form [4]:

$$x = p^v \frac{m}{n}$$

with n and m not divisible by p. The p-adic norm of the rational number is equal to:

$$|x|_p = \frac{1}{p^v}$$

The field of p-adic numbers $Q_p$ is the completion of the field of rational numbers Q with the p-adic norm . The most interesting property of p-adic numbers is their ultrametricity. This means that they obey to the strong triangle inequality:

$$|x+y|_p \leq \max(|x|_p, |y|_p)$$

Let us remind that a real number may be expressed by following the expansion:

$$10^v \sum_{n=0}^{\infty} b_n \left(\frac{1}{10}\right)^n$$

here $b_n = (0, 1, ...., p-1)$. A p-adic number has the following expansion:

$$x = p^v \sum_{n=0}^{\infty} a_n p^n$$

Here $a_n = (0, 1, ...., p-1)$. Furthermore we can define addition, subtraction, multiplication and division operation. Today there exist the algebra and analysis on the field of p-adic numbers.
Let consider the free p-adic theory which gives the following formal solution is x=Ct+B, and C, B p-adic constants. This is the geodesics of the free theory. To obtain the final result we need to construct some type of mapping from p-adic numbers to real numbers. This will give us the opportunity to compare our results with the price dynamics. Let us take the following form of the mapping:

$$a_n \to (a_n)^D$$

A parameter D is called the dimension of the fractal space. The readers can learn the situation in this sphere in [1]. In the charts below two different kinds of waves with comparison against the real data – sawlike and steplike waves are shown.
It is seen that p-adic function can be very effective for the interpolation of this type of a signals

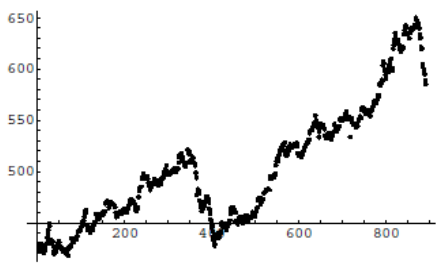 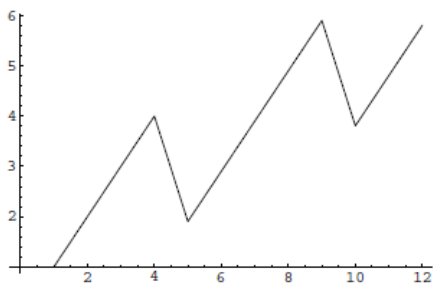

Fig 2. *Russian stock Index*           Fig.3. *Subcritical wave (First Level of Fractal) for D>1, p=3*

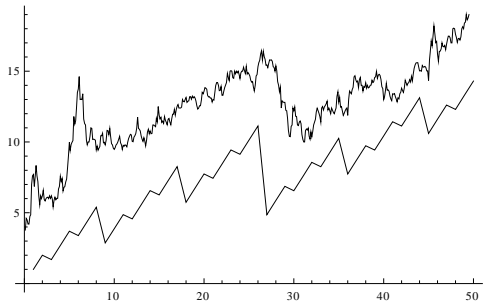
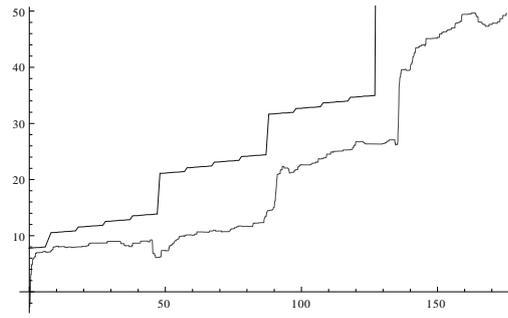

Fig 4. *Subcritical wave (Third Level of Fractal)for D>1, p=3  The second curve shows the real data*.

Fig 5. *Supercritcal wave (third Level of Fractal) for D<1, p=3
This type of wave is not presented in the Elliott theory.*

One can see that this is the basic elements of Eliott wave analysis which is well known among the financial analysists. We have here two type of the waves: (1) is the supercritical wave and (2)-is the subcritical one.  So, we see say that the p-adic and the adelic theory give us the foundation  of Eliott type theory.

The figure below shows that the graphs of stock market crashes very similar to the simple p-adic configuration , Sornette [5]

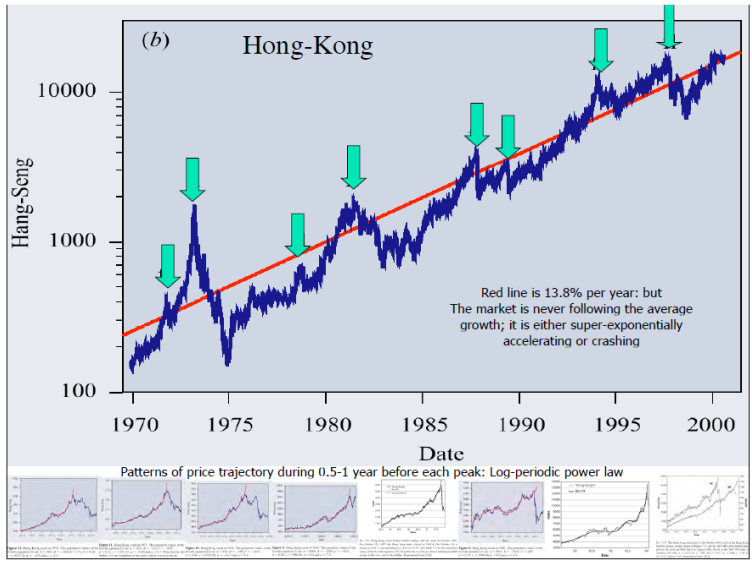

The figures of Gazprom, IBM  shares and the RTS Index are shown below. The second curve shows the procedure a p-adic interpolation of real data. Different time scales are used. The second curve shows the real data.

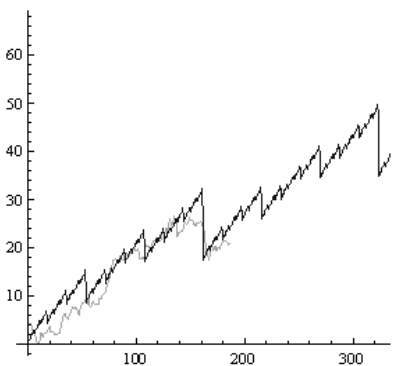
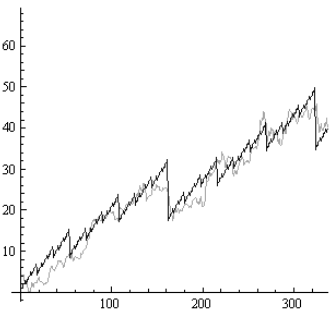

*IBM Year timeframe*

```
        01.07.2006-01.04.2007              01.07.2006-01.07.2008
```
In first figure interpolation of real data was shown. Forecast of future value was placed(PROGNOZ!). On second figure this forecast was compared with real data.

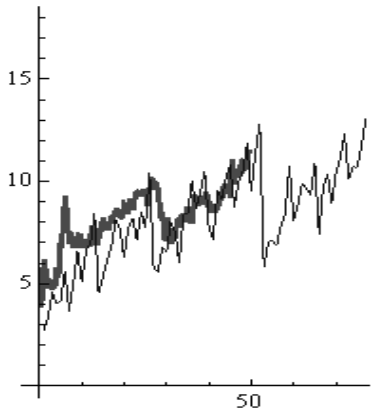 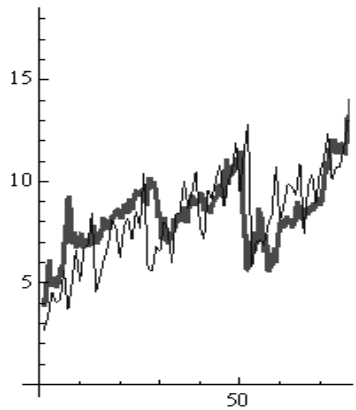

*Gazprom Daily time frame*
```
      01.06.2009                    01.06.2009-02.06.2009
```

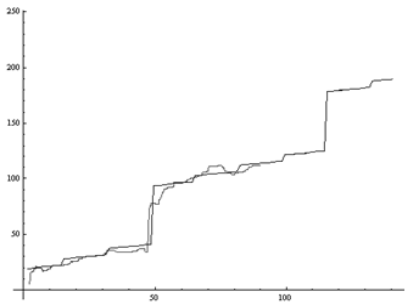 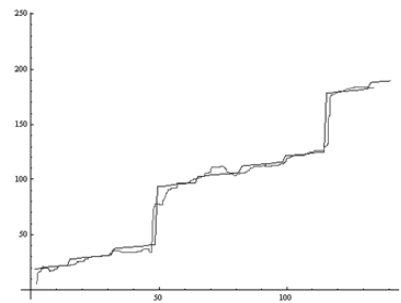

*RTS index Weekly time frame*
```
      27.05.2009-30.05.2009         27.05.2009-1.06.2009
```

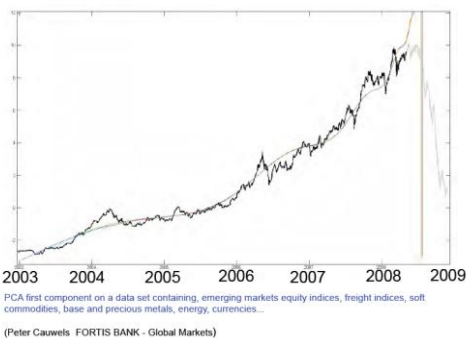 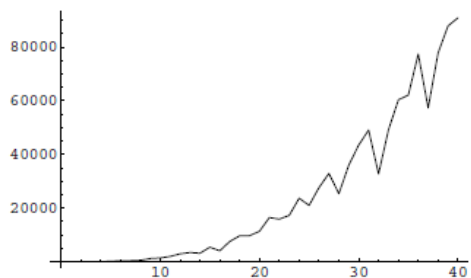

   Sornette   Theory (Log Periodic)          P-adic Theory (Power Low Function Only)

This Figure shows the power low function $x^3$ with D=0.45.

In this figure comparison was made of typical bubble [5] with p-adic $x^3$ function with fractal dimension 0.45. Thus, our hypothesis is that the financial market is described by the p-adic numbers and maps, was confirmed. P-adic mathematics provides a good mathematical framework to describe the Elliott Wave. The use of even simple database (p = 2, 3) for a fractal approximation possible to qualitatively describe the Elliott wave patterns and other so-called "ladder", which in the theory of waves are not described, but rather common in the Russian stock market. Application of p-adic approach to the small number of parameters simplifies and accelerates the approximation  of wave patterns in financial markets, but also allows extrapolation of the price trend.

## 4. Adele

Adele a is the set of the following type [4]:

$$a = (a_\infty, a_2, \ldots, a_j, \ldots),$$

As for us it will be very important that an adele group has an additive character (the analog of exponential function of plane wave):

$$\chi(xy) = \chi_\infty(x_\infty y_\infty) \prod_p \chi_p(x_p y_p) = \exp(-2\pi i x_\infty y_\infty) \prod_p \exp(2\pi i \{x_p y_p\}_p),$$

is the fractional part of $x_p y_p$. It is evident that an adele plane wave contains the infinite numbers of the different p-adic projections of it. A multiplicative character of an adele group has the following expression:

$$\pi(b) = \pi_\infty(b_\infty) \pi_2(b_2) \ldots \pi_p(b_p) = |b_\infty|_\infty^s \prod_p |b|_p^s = |b|^s,$$

The price is described by superposition of elementary Bruhat-Swarts functions:

$$\varphi(x) = \varphi_\infty(x_\infty) \prod_p \varphi_p(x_p);$$
$$x \in A^+; \; \varphi_\infty(x_\infty) \in S(R); \varphi_p(x_p) \in S(Q^p).$$

As will be shown later that an each p-adic component of this function describes some fractal regime. That is why all components of this function describes in its turn the multifractal behavior of the price. We restricted the set of adele function components from the following set of function:

$$(\varphi_\infty, \varphi_2, \varphi_3).$$

The rest components of adele function will be equal to a function:

$$\varphi_p(x_p) = \Omega(|x|_p).$$

In the microscopic theory there exist three distinct regimes of describing the activity of traders. On the traders level we have the theory with a set of fields which depend on the decision making variables of an each individual trader. It is a high energy regime (analog of the UV regime in the field theory). At the medium range of energy there appears a some p-adic formulation as a result of the spontaneous breaking of some symmetry. It gives us a description of trends and minitrends of prices, and it is the an analog of fractal description of price behavior. In the low energy limit (IR regime in the field theory) there appears the adelic (multifractal) description of the market.

## 5. Adele functional integral

All these above mentioned regimes are described by three different types of quantum mechanics. According to Vladimirov and Volovich we have the following possibilities: the usual quantum mechanical formalism describes the microscopic mechanisms, which exist between traders; the p-adic quantum mechanical formalism describes the typical patterns behavior of prices and the last one - an adelic quantum mechanical formalism corresponds to the total dynamics of the securities price. These types of quantum description are presented by the following triplet of objects:

usual quantum mechanics: $(L_2(R), W(z_\infty), U(t_\infty)),$

p-adic quantum mechanics: $(L_2(Q_p), W_p(z), U_p(t))$,

adelic quantum mechanics: $(L_2(A), W_p(z), U_p(t))$,

here $Q_p$--p-adic number field, z=q,x- p-adic coordinates and momentums, $L_2(Q_p)$ -space of square integrable functions in Hilbert space of the system;

$W_p(z)$ - -an unitary representation of Heisenberg-Weil group; $U_p(t)$ - -an evolutionary operator. W - operator gives the Weil representation for the commutation relations and acts according to the integral operator:

$$W_p(z)\psi_p(x) = \int_{Q_p} W_p(z; x, y)\psi_p(y)dy,$$

where $\psi_p \in L_2(Q_p)$ with the following integral kernel:

$$W_p(z; x, y) = \chi_p(2kx + qz) * \delta(x - y + q).$$

An evolutionary operator is given by the following integral kernel. In the adelic quantum mechanics the state of a system is given by the function:

$$\Psi_c(x, t) = \Psi_\infty(x_\infty, t_\infty)\prod_p \Psi_p(x_p, t_p) \prod_{p \in c} \Omega(|x|_p).$$

An adelic evolution operator has the infinite numbers of p-adic components:

$$U(x, t) = \prod_p U_p(x_p, t_p)$$

The action of concrete component is defined by the integer operator:

$$U_p(x_p, t_p)\psi_p(x) = \int_{Q_p} K_p^t(x, y)\psi_p(y)dy;$$

the kernel $K_p(x_p'', t_p''; x_p', t_p')$ is defined through the functional integral:

$$K_p(x_p'', t_p''; x_p', t_p') = \int \chi_p(-S[x])Dx =$$

$$\int \chi_p\left(-\int_{t'}^{t''} L[x_m, t_m, x'_m]dt_m\right)\prod_{t_m} dx(t_m).$$

## 6. Minority Game

Minority game is described by the following diagram:

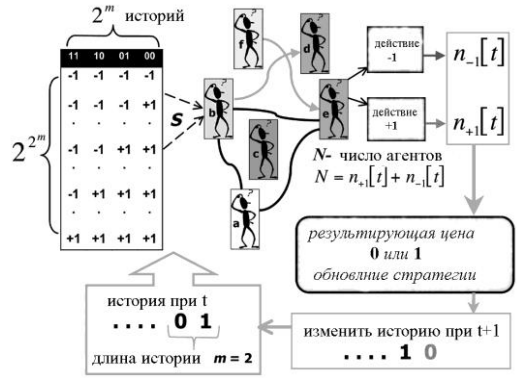

During the last time some version of the microscopic stock market description were presented in many papers. The most famous of them is the so called minority game model, which was given by Zhang and Challet [6]. It is the agent base model of stock market. In this model, the traders execute operations (buy and sell), which can be described by spin variables $s_i$. The minority game involves N traders, labeled with Roman indices i,j,k, etc. At each round of the game, all the traders act on the basis of the one and the same external information I(k). Each trader i has at his disposal S number of strategies. The volatility of the market is described by the expression:

$$\sigma^2 = \sum_\mu (A^\mu)^2 - (\sum_\mu A^\mu)^2,$$

where $A^\mu = N^{-\frac{1}{2}} \sum_i [\omega_i^\mu + s_i \xi_i^\mu]$ is the action of all traders. Here from:

$$\sum_\mu A^\mu = \frac{1}{N\sqrt{N}} \sum_i s_i (\sum_\mu \xi_i^\mu),$$

$$\sum_\mu (A^\mu)^2 = \frac{1}{2} + \frac{1}{N} [\sum_i h_i s_i + \frac{1}{2} \sum_{ij} J_{ij} s_i s_j]$$

where $J_{ij} = \xi_i^\mu \xi_j^\mu$ -are some coefficients describing the traders strategies and $\mu$ =(1,2,3,...K) -variable, describing the history [6]. The second expression formally coincides with the hamiltonian of a spin glass.

## 7. Hubbard-type microscopic model

At present time it is very desirable to generalize this model, to make it more adequate to traders activities and to take into account three or four states of traders (buy, sell, hold and ground state). It is well known in condensed matter theory that spin glass model or Heisenberg model is a "square root" of Hubbard model. It is natural to use here a Hubbard type model:

$$H = -t \sum_{<r,r'>,s} \alpha_{r,s}^+ \alpha_{r',s} + U \sum_r n_{r,\uparrow} n_{r,\downarrow} =$$

$$\sum_r U X_r^{22} + \sum_{A,C,r,r'} t_{-AC}(r-r') X_r^{-A} X_{r'}^{C}$$

where $<r, r'>$ -denote the sum over the nearest neighbors and $r$ parameterise trader i . Traders are described by $(\alpha_{rs}^+, \alpha_{rs})$ –creation and destruction operators , s - the spin (gives the decision making variable) of traders. We give two different forms of this model; the first form is a standard one, the second form contains the Hubbard operators $X^A$ (in fact they are projectors $X_r^A = |pr\rangle\langle qr|$). These operators act in space of following states: $|0\rangle$ -is the ground state of a trader, $|\uparrow\rangle = \alpha_\uparrow^+ |0\rangle$ -is the buy state of a trader, $|\downarrow\rangle = \alpha_\downarrow^+ |0\rangle$ -is the sell state of a trader, $|2\rangle = \alpha_\uparrow^+ \alpha_\downarrow^+ |0\rangle$ -is the hold state of a trader. Such types of states of the traders appear in paper of Thomas Lux in his theory of stock market [7] Here the first term describes the trading activity: buying of an i r trader and selling of an j ( r) trader; the second term describes the distribution of the capital among the traders. These models gives us the description the microscopic picture of trading. This formulation contains some variables which are determined by strategies of the traders. After integration over these variables in the functional integral we have obtained some effective theory. But the theory which describes price dynamics as the result of collective behavior of ensemble of traders can be derived from the previous theory by the application of the generalized supercoherent state. Using this way we have obtained an effective functional formulation.

$$L_{eff} = \frac{\langle G(\theta,\mathbf{r},t') | (\frac{\partial}{\partial t'} - H) | G(\theta,\mathbf{r},t') \rangle}{\langle G(\theta,\mathbf{r},t') | G(\theta,\mathbf{r},t') \rangle},$$

where $|G\rangle$ -is a supercoherent state, which is expressed through generators of the dynamic superalgebra; $\{\mathbf{r},t',\theta\}$ –supercoordinates of superspace.

$|G\rangle$ can be constructed by following way [8]:

$$|G\rangle = e^{-\sum_{k=1}^{6} X^k b^k(\mathbf{r},t,\theta) - \sum_{j=1}^{2} X^{-j} \chi^j(\mathbf{r},t,\theta)} |0\rangle$$

here $|0\rangle = \otimes_r |0\rangle_r$, $\{b^C\} = \{\{E_i\},\{h_i\}\}$ $i = 1,2,3$ and $\{X_{05}, X_{50}, X_{00} - X_{55}, X_{\downarrow\uparrow}, X_{\uparrow\downarrow}, X_{\downarrow\downarrow} - X_{\uparrow\uparrow}\}$ $\{X_1, X_3\} = \{X_{0\downarrow} + X_{\uparrow 5}, X_{0\uparrow} - \dot{X}^{\uparrow 2}\}$ – set of the bosonic fields (even valued grassmanian fields). $|G\rangle$ has four component, two of them are fermionic (odd valued grassmanian nonlinear composite fields), and two are bosonic (also composite and nonlinear in $\chi, E, h$) [9]. This theory as shown in recent paper gives the p-adic functional integral and description and can be regarded as microscopic model of market. Let us describe the possible scenario of this functional integral investigation. We have the very nonlinear representation which contain quantum group. This quantum group formulation can be transformed to so called q-analysis. When q=1/p we have an p-adic representation for our functional integral. This p-adic regime was described on the top of this article.

## 8. Conclusion

The main conclusion of this work consists in the statement that the price is an adelic function. We formulate in this article the deep program of investigation of the microscopic theory of stock market.